\documentclass[a4paper]{amsart} 
\usepackage[frenchb,english]{babel} 
\usepackage[latin1]{inputenc} 
\usepackage{enumerate, eufrak, amsmath} 
 
\def\ind{{\mathchoice {\rm 1\mskip-4mu l} {\rm 1\mskip-4mu l} 
{\rm 1\mskip-4.5mu l} {\rm 1\mskip-5mu l}}} 
\newcommand{\e}{\mathrm{e}}

\begin{document} 
\title{Remarks on the statistical study of protein-protein interaction in 
  living cells}  
 
\author{Ph. Heinrich \and J. Kahn}
\address{Laboratoire Paul Painlev\'e\\
UMR CNRS 8524\\ 
Universit\'e Lille 1\\ 
Cit\'e Scientifique \\ 
59655 Villeneuve d'Ascq Cedex\\ 
France} 
\email{philippe.heinrich@univ-lille1.fr, jonas.kahn@math.univ-lille1.fr}

\author{L. Héliot \and D. Trinel}
\address{Interdisciplinary Research Institute\\ 
Parc de la Haute Borne\\ 
50 avenue de Halley 
BP 70478\\ 
59658 Villeneuve d'Ascq Cedex\\ 
France} 
\email{laurent.heliot@iri.univ-lille1.fr, dave.trinel@iri.univ-lille1.fr} 
 

\keywords{Maximum likelihood, multi-exponential model, model selection, FRET, FLIM, TCSPC} 
   
 \subjclass[2010]{62F03, 92C40} 
 
\selectlanguage{english}
\begin{abstract} In this note, we focus on a selection model problem: a 
  mono-exponential model versus a bi-exponential one. This is done in the 
  biological context of living cells, where small data are 
  available. Classical statistics are revisited to improve existing results. 
Some unavoidable limits are also pointed out.  
\end{abstract} 
 
\maketitle 
\section{Introduction} 
 
The measurement of molecular dynamic interactions and their respective 
proportions in living cells or tissues is a major question in 
biological and medicine research. The Förster resonance energy transfer
(FRET) is one of the best known approaches to observe and 
quantitatively study protein-protein interactions at a subcellular 
level (\cite{STRV}). The FRET measurement can be currently performed by fluorescence lifetimes imaging 
microscopy (FLIM for short) in living cells and tissus. It can be achieved via the time correlated single photon 
counting (TCSPC) method which provides a lifetime decay curve per 
site (\cite{WSH}).  To be interpreted, this curve is fitted by selecting the 
``best'' (with respect to a given statistical criterion) 
multi-exponential model. Contrary to a mono-exponential model, a 
bi-exponential one witnesses interaction between two proteins. Our aim is to 
find, pixel per pixel, which of these models is accurate. But 
one difficulty is that the number of observed photons per pixel is small for any statistical treatment in order to preserve the living cell and therefore cannot be 
increased. An attempt to deal with the problem can be found in \cite{STRV}. Our 
aim here is to go further in this direction pointing out some improvements and  
limits. Some account of statistical methods in this area can be found in 
\cite{HS} and \cite{MM}. 
 
\subsection{Modelling fluorescence lifetimes} 
 
It is not necessary to describe here in details FLIM and TCSCP. We only need 
to understand that lifetimes are measured as differences between excitation 
times (pulses) and emission times of photons. Denote by $r$ the period between 
two consecutive pulses. Here $r$ is $12$ nanoseconds, near values taken in practice. What is actually 
measured is a lifetime modulo $r$ since we cannot be sure from what pulse it 
goes. 
 
It is assumed that lifetimes come from say $K$ species and are observed in the 
interval $[0,r)$ after infinitely many pulses. In these conditions, each 
lifetime species $k$ ($1\le k\le K$) admits the following probability density: 
\begin{equation} 
  \label{eq:densityk} 
 f_k(t)=\alpha_k\exp(-\alpha_kt)\frac{\ind_{[0,r)}(t)}{1-\exp(-\alpha_kr)}  
\end{equation} 
where $\alpha_k$ is the inverse mean lifetime of the $k$-th species. A uniform 
noise is added with density 
\begin{equation} 
  \label{eq:bruit} 
 f_0(t)=\frac{\ind_{[0,r)}(t)}{r}.   
\end{equation} 
If $\pi_k$ denotes the proportion of the $k$-th species ($\pi_0$ refers to the 
noise's one), we get the probability density of the fluorescence lifetime by 
writing  
\begin{equation} 
  \label{eq:density} 
 g(t)=\sum_{k=0}^K \pi_kf_k(t).  
\end{equation} 
 
\subsection{Modelling the photon emission} 
Let $I_k$ be the mean photon number of species $k$ detected between two 
pulses. Assume that photons occurrences are independent. Then the total number 
of detected photons is Poisson distributed with intensity $T\sum_{k=0}^KI_k$ if 
observations take place during $T$ pulses. For a later use, it is 
convenient to set  
\begin{equation} 
  \label{intensite} 
I=\sum_{k=0}^KI_k.  
\end{equation} 
Note that we have  
$$\pi_k=\frac{I_k}{I}.$$  
Since the noise intensity $I_0$ will be supposed known, it is convenient to 
consider proportions $\pi'_k$  among all  species with $k\ge 1$ except 
$k=0$. Thus, we have for $k\ge 1$, 
$$\pi_k'=\frac{I_k}{I-I_0}=\frac{I}{I-I_0}\,\pi_k.$$ 
 
\subsection{Maximum likelihood estimation (MLE) and likelihood ratio test} 
The aim is firstly the determination of the most probable parameter 
$\theta:=(\alpha_1,\ldots,\alpha_K,I_1,\ldots,I_K)$ from observed lifetimes 
modulo $r$ denoted by $t_1,\ldots,t_n$. The noise intensity $I_0$ is supposed 
known. The related log-likelihood is then 
\begin{equation} 
  \label{eq:vrais} 
\mathfrak{L}(\theta)=\mathfrak{L}(\theta ;t_1,\ldots,t_n)=\\-IT+n\log(IT)-\log(n!)+\sum_{i=1}^n\log\left(g(t_i)\right). 
\end{equation} 
For physical reasons, in particular since lifetimes are sure to be between
30pc and 30ns, we may and do assume that $\theta$ lies in a compact parameter
set.

Numerical optimisation of the likelihood \eqref{eq:vrais} is made easier by knowing derivatives: 
\begin{eqnarray*} 
  \label{eq:dll}  
 \frac{\partial g(t)}{\partial I_k} & = & \sum_{l=0}^K 
\frac{I_{l}}{I^2}\left[f_k(t)-f_{l}(t)\right]=\frac{f_k(t)-g(t)}{I}; \\ 
\frac{\partial g(t)}{\partial \alpha_k} & = &\ind_{[0,r)}(t) \frac{I_k}{I}\frac{\e^{-\alpha_kt}}{\left(1-\e^{-\alpha_kr}\right)^2}\left(1-\alpha_kt+(\alpha_kt-\alpha_kr-1)\e^{-\alpha_kr}\right);\\ 
 \frac{\partial \mathfrak{L}(\theta)}{\partial I_k} & = & -T +\frac{n}{I}+\sum_{i=1}^n\frac{f_k(t_i)-g(t_i) }{Ig(t_i)}=-T+\frac{1}{I}\sum_{i=1}^n\frac{f_k(t_i)}{g(t_i)};\\ 
\frac{\partial \mathfrak{L}(\theta)}{\partial \alpha_k} & = &\sum_{i=1}^n\frac{\frac{\partial g(t_i)}{\partial \alpha_k}}{g(t_i)}=\frac{I_k}{I}\sum_{i=1}^n\frac{\frac{\partial f_k(t_i)}{\partial \alpha_k}}{g(t_i)}. 
\end{eqnarray*} 
Denote by $\theta_K^*$ the most probable parameter if there is $K$ species.  
 
To decide next 
which model from $K=1$ or $K=2$ is the most accurate, a classical 
statistic is the likelihood ratio  
$$D:=\left[\mathfrak{L}(\theta_2^*)- \mathfrak{L}(\theta_1^*)\right].$$ 
From a theoretical point of view, since we are dealing  with the number of
components of a mixture model, even the asymptotics under the null hypothesis
are not the usual $\chi^2$ statistics. It can be expressed as a supremum over
a Gaussian process on a subset of a four-dimensional unit sphere (in our case) endowed with the ``right'' covariance function (\cite{Gas2} and references therein). However this process depends also on the ``true'' point $\theta $. Since all calculations are complicated, it is easier to simply simulate if we want to know the level of a test associated with a given threshold. 
 
Notice on the other hand that simulations hint that the likelihood ratio test
is quite efficient for knowing the number of components in a mixture with
compact parameter set (see for
example \cite{Gof} or \cite{Chen1}).

\section{Selection of the number of exponential species $K$} 
\subsection{Comparisons} 
We restricted ourselves to test $K=1$ versus $K=2$. It can be already a 
difficult and interesting question, if few observed photons are available. With the 
help of simulated observations, we first optimised $\theta$ 
by MLE for each $K$ and next tested $K=1$ versus $K=2$ via the likelihood 
ratio statistics $D$. 
 
Compared to the one given in \cite{STRV}, the preceding statistical test is as 
efficient but with about 100 times less observations. For the reader's 
convenience and for comparison, consider the table obtained in \cite{STRV}: 
\begin{table}[!htp] 
  \centering 
  \begin{tabular}[c]{|c|c|c|c|c|c|c||c|} 
 \hline 
Nbr of photons $/\Delta\chi^2$ & 10.0 &20.0& 30.0&40.0& 50.0& 90.0& Error(\%)\\\hline 
1000 & 35.7 & 34.8 & 34.3&34.6&34.9 & 45 & $>$ 20\\ 
10000 & 13.7 & 12.0 & 11.9 & 12.1& 12.9 & 27.3 & $<$ 20\\ 
100000 & 4.2& 1.7 & 2.3 & 2.7 & 4.7& 26.3& $<$ 5\\ 
1000000 & 1.7 & 0.0 & 0.0 & 0.0 & 3.3 & 32.7& $<$ 2\\ 
\hline    
  \end{tabular} 
\medskip\\ 
  \caption{Frequency of selection of the wrong model. It depends on the 
    observations number and a $\Delta\chi^2$ criterion which consists in 
    comparing the $\chi^2$ statistics for $K=1$ and $K=2$. Simulations were 
    performed on a mix of $1/\alpha_1=0.6$ ns and $1/\alpha_2=2.4$ ns with 
  different proportions $\pi'_1=0,.077,.2,.43,1$ with $100$ noise photons. 30 
  simulations per condition.} 
\end{table} 
 
In similar simulation conditions, we have obtained the following: 
 
\begin{table}[!htp] 
  \centering 
  \begin{tabular}[c]{|c|c|c|c|}  
 \hline 
Nbr of photons & Mean error rate (\%) &  Best threshold  & Mean error rate \\ 
 & & & at threshold 4 \\\hline 
1000 & 12.8 & .85 & 20 \\ 
10000 & 0.3 & 4 & 0.3 \\  
100000 & 0 &  4 & 0 \\  
\hline     
  \end{tabular}  
\medskip\\  
  \caption{Frequency of selection of the wrong model. It depends on the  
    observations number and a likelihood ratio criterion. Simulations were  
    performed on a mix of $1/\alpha_1=0.6$ ns and $1/\alpha_2=2.4$ ns with  
  different proportions $\pi'_1=0,.077,.2,.43,1$ with $100$ noise photons. 500 simulations for each proportion and number of photons.}  
\end{table}  
Here, mean error rate is the average over the simulation number of the 
percentage to select the wrong model. Best threshold means threshold that 
gives the smallest mean error rate; using a very crude optimisation.  
Notice that the strange values $0,.077,.2,.43,1$ of $\pi_1'$'s proportion 
correspond to values $.25, .5, .75,1$ of proportion $\eta_1=\dfrac{\pi'_1\alpha_2}{\pi'_1\alpha_2+\pi'_2\alpha_1}$ considered in \cite{STRV}.   
A consequence is that we never test more short-life photons than 
long-life. Moreover the case $\pi_1' = .077$ is not very far away from the 
mono-exponential case. In particular, with $1000$ photons, among which $100$ 
noise photons, the expected number of photons with $0.6$ nanoseconds lifetime 
is less than the number of noise photons. If we compute the error rate for 
$1000$ photons without that case, we obtain for instance a $2.6\%$ error rate 
for the likelihood ratio test at threshold $3$.  
 
\subsection{Simulation scheme} 
 
The data set generation algorithm is as follows: 
\begin{enumerate}[1.] 
\item Sample $n_k$ the number of photons for each species $k$, including 
noise ($k=0$), from a Poisson distribution of parameter $TI_k$. 
\item Draw $n_k$ lifetimes with distribution density $f_k$ for each species $k$. 
\item Return the set of all the sampled lifetimes, regardless of $k$. 
\end{enumerate}

Some differences between simulation methods should be noted:  
\begin{itemize}  
\item We use random Poissonian number of photons rather than fixed number of  
photons : we take into account the ``offset noise''. 
\item Instrumental response: we neglect the $.03$ nanoseconds long 
  instrumental response function. 
\item Exact times \emph{vs} channels: we did not use bins and worked as if we 
  knew the exact detection times. 
\end{itemize} 
Nevertheless these differences should have little effect and comparisons still 
make sense.  
 
\section{Further comments} 
\subsection{With closer lifetimes} 
If we choose $1/\alpha_1=1$ ns and $1/\alpha_2=2$ ns as mean lifetimes, it is harder to select the right number of species:  
\begin{itemize} 
\item With 10000 photons and  $\dfrac{\pi_0}{1-\pi_0}=.01$ as noise ratio: 
  \begin{itemize} 
  \item If $\pi_1'=\pi'_2=.5$ or  $\pi_1'=.75,\,\pi'_2=.25$, 
  no wrong selection should occur, 
\item If  $\pi_1'=.25,\,\pi'_2=.75$, the error rate is about $.1\%$ when the 
  threshold is calibrated so as to balance errors ``mono towards bi'' and ``bi 
  towards mono''.   
  \end{itemize} 
\item With 1000 photons and  $\dfrac{\pi_0}{1-\pi_0}=.01$ as noise ratio: if 
  $\pi_1'=\pi'_2=.5$, the error rate is about $15\%$ when the 
  threshold is calibrated so as to balance errors ``mono towards bi'' and ``bi 
  towards mono''. 
\end{itemize} 
If we choose close mean lifetimes such as  $1/\alpha_1=1.4$ ns and 
$1/\alpha_2=1.6$ ns, we are too close to the ``border'' of the model, and about 1 
million photons is required to distinguish the two components. By border, we 
mean a proportion close to $0$ or $1/\alpha_1$ close to $1/\alpha_2$ so that 
identifiability problems occur with small samples. Asymptotically, when we get $n$ times closer to the border of a mixture model, we need $n^{4}$ times as many photons to get the same statistical efficiency, for any procedure \cite{JiahuaChen}. 
 
\subsection{Absolute limits} 
\label{abslim} 
 
The former sentence about rates when we get nearer the border is a first expression of limits that cannot be broken, no matter how smart the statistical procedure. To give a small taste of what to expect, here are the best error rate when having to choose specifically between two possible sets of lifetime parameters and corresponding distribution probabilities $f^1$ and $f^2$, with equal \emph{a priori} probabilities. In that situation, which is easier than the one studied in the article, the optimal choice is the one with greater observed likelihood, and the error rate is $\frac12 - \frac{1}{4} \left\lVert f^1 - f^2 \right\rVert_1  $. 
\begin{itemize} 
\item{With $32$ observed photons and a signal to noise ratio of $1/10$, choose between a mono-exponential with lifetime $2.4$ ns, and a bi-exponential with proportions $0.077$ and $0.923$ and lifetimes $0.6$ and $2.4$ ns: optimal error rate $>25\%$.} 
\item{With $32$ observed photons and no noise, choose between a mono-exponential with lifetime $2.6$ ns, and a bi-exponential with proportions one half and lifetimes $2.5$ and $2.7$ ns: optimal error rate $>49.75\%$.} 
\end{itemize} 
 
The second case is almost as bad as a coin toss, ignoring the data.

\bibliographystyle{plain}

\bibliography{bibliodjp}

\end{document}